\documentclass[twoside]{ilcws08}
\usepackage[latin1]{inputenc}
\usepackage[dvips]{graphicx,epsfig,color}
\usepackage{wrapfig,rotating}
\usepackage{amssymb,amsmath,array}

\begin{document}
\title{
Dominant NNLO Corrections to Four-Fermion Production at the $WW$ Threshold} 
\author{Stefano Actis
\thanks{Work supported by the DFG through SFB$/$TR9. 
        Report numbers: PITHA 09$/$03; SFB$/$CPP-09-07.
        Slides available at~\cite{url}.}
\vspace{.3cm}\\
Institut f\"ur Theoretische Physik E, RWTH Aachen University, \\
D-52056 Aachen - Germany
}

\maketitle

\begin{abstract}
  The recent evaluation of the parametrically dominant 
  next-to-next-to-leading order corrections to four-fermion 
  production near the $W$-pair threshold in the framework 
  of unstable-particle effective theory is briefly summarized.
\end{abstract}

\section{Introduction}

The production of $W$-boson pairs at electron-positron colliders 
is a process of crucial relevance for a precise determination 
of the $W$ mass. If the International Linear Collider will measure 
the total cross section at the per-mille level~\cite{AguilarSaavedra:2001rg}, 
a direct reconstruction of the $W$-decay products will allow 
to reach a $10$ MeV accuracy on the determination of the $W$ 
mass~\cite{monig}. A higher precision could be achieved through a 
dedicated threshold scan leading to a $6$ MeV accuracy~\cite{wilson}.

The aforementioned estimates rely on statistics and the performance 
of the future collider, and assume that the cross section for $W$-pair 
production is theoretically under control. In particular, in view of the 
$6$ MeV precision goal, accurate predictions are needed for a final 
state containing the fermion pairs produced by $W$ decay, instead of 
on-shell $W$ bosons.

A full next-to-leading order (NLO) evaluation of four-fermion 
production in the complex-mass scheme has been performed by 
the authors of~\cite{Denner:2005es}, extending the methods introduced 
in ~\cite{Denner:1999gp}. Recently, a compact analytic result for the 
threshold region has been derived in ~\cite{Beneke:2007zg} 
(see~\cite{Schwinn:2007mv} for reviews) using the method of 
unstable-particle effective field theory~\cite{Beneke:2003xh}.

The work of~\cite{Beneke:2007zg} has concluded that collinear 
logarithms arising from initial-state radiation have to be re-summed 
at next-to-leading accuracy for reducing  the threshold-scan error 
on the $W$ mass to less than $30$ MeV. Furthermore, it has been 
shown that the NLO partonic evaluation in the effective-theory 
framework is affected by a residual error of $10-15$ MeV. Although 
a large part of the uncertainty at the partonic level can be removed 
using the full NLO result of ~\cite{Denner:2005es}, the evaluation 
of the dominant next-to-next-to-leading order (NNLO) corrections 
is mandatory to secure the $6$ MeV threshold-scan accuracy goal.

In~\cite{Actis:2008rb} we have evaluated the parametrically 
dominant NNLO corrections to the total cross section for the 
production process $e^- e^+ \to \mu^- \overline{\nu}_\mu u 
\overline{d} + X$, where $X$ is an arbitrary flavor-singlet state. 
The result is expressed through a compact semi-analytic formula 
that can be easily added on top of both 
effective-theory~\cite{Beneke:2007zg} and full NLO~\cite{Denner:2005es}
predictions. 

In Section~\ref{sec:NNLO} of this note we show an overview 
of the NNLO corrections. Next, in Section~\ref{sec:num}, we
discuss their numerical impact.

\section{Overview of the dominant NNLO corrections}
\label{sec:NNLO}

The inclusive cross section for the process $e^- e^+ \to 
\mu^- \overline{\nu}_\mu u \overline{d} + X$ is computed 
in the context of the effective theory~\cite{Beneke:2003xh} 
by means of a non-standard perturbative expansion in three 
small parameters of the same order $\delta$:
1) $\alpha_{ew} \equiv \alpha \slash \sin^2 \theta_w $, where 
   $\alpha$ is the fine-structure constant and $\theta_w$ stands 
   for the weak-mixing angle; 
2) $(s - 4 M_W^2) \slash (4 M_W^2) \sim v^2$, where $s \equiv 
   (p_{e^-} + p_{e^+})^2$, $M_W$ is the $W$ mass and $v$ is the 
   non-relativistic velocity of the $W$;
3) $\Gamma_W \slash M_W$, with $\Gamma_W$ denoting the $W$ decay
   width.

The re-organized loop and kinematical expansion is performed 
through the method of regions~\cite{Beneke:1997zp} and relies 
on the identification of different momentum scalings in the 
center-of-mass frame in order to exploit the hierarchy of scales 
around threshold. Denoting by $k$ an arbitrary loop-integration 
momentum, we deal with hard ($k_0\sim|\vec{k}|\sim M_W$), potential 
($k_0\sim M_W \delta, |\vec{k}| \sim M_W \sqrt{\delta}$), soft 
($k_0\sim |\vec{k}| \sim M_W \delta$), collinear ($k_0 \sim M_W, 
k^2 \sim M_W^2 \delta$) and semi-soft ($k_0\sim |\vec{k}|\sim M_W 
\sqrt{\delta}$) momentum scalings. Semi-soft modes are not relevant 
for the NLO evaluation~\cite{Beneke:2007zg}, and start playing a 
role for the NNLO calculation~\cite{Actis:2008rb}. 

After integrating hard modes out, the residual dynamical degrees 
of freedom contribute to genuine loop computations in the context 
of the effective theory. The different scaling properties lead to a
peculiar half-integer power counting in the expansion parameter 
$\delta$ and to a straightforward identification of the parametrically
dominant radiative corrections.

\begin{wrapfigure}{r}{0.7\columnwidth}
  \centerline{\includegraphics[width=0.65\columnwidth]{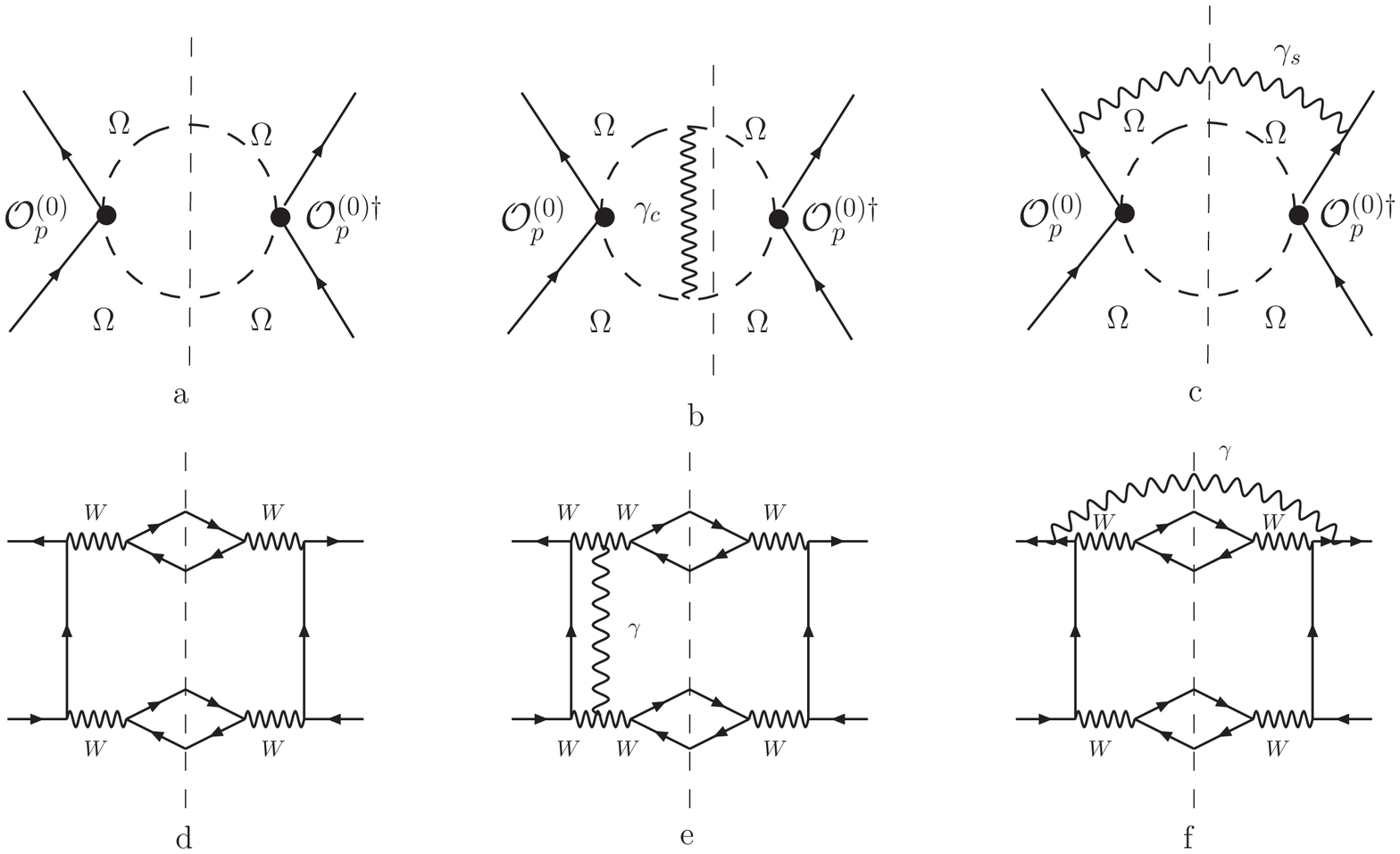}}
  \caption{sample LO and NLO diagrams in the effective theory 
           (first line) and in the full Standard Model (second line).}
  \label{Fig:Cuts2}
\end{wrapfigure}

The total cross section for four-fermion production is computed 
from the cuts of the $e^- e^+$ forward-scattering amplitude, as 
shown in Figure~\ref{Fig:Cuts2}a for the leading order (LO) 
diagram (see Figure~\ref{Fig:Cuts2}d for the Standard Model 
counterpart). Here the LO operator ${\cal O}_p^{(0)}$ 
(${\cal O}_p^{(0)\dagger}$) accounts for the production (destruction) 
of a pair of non-relativistic $W$ bosons, denoted by $\Omega$. In 
Figure~\ref{Fig:Cuts2}b and Figure~\ref{Fig:Cuts2}c we show also 
the NLO Coulomb- and soft-photon corrections  evaluated 
in~\cite{Beneke:2007zg}. The conventional  Standard Model (SM) 
loop expansion of Figure~\ref{Fig:Cuts2}e and Figure~\ref{Fig:Cuts2}f 
treats virtual Coulomb effects ($\gamma_c$) and soft real-photon 
contributions ($\gamma_s$) as genuine NLO terms. In the framework 
of the effective theory, instead, a simple power-counting argument 
shows that Coulomb corrections at the $W$-pair threshold are 
suppressed by a factor $\delta^{1\slash 2}$ with respect to the 
LO result, and can be classified as dominant NLO effects, whereas 
soft-photon diagrams, being weighted by one power of $\delta$,
lead to sub-dominant NLO effects. 

Relying on analogous observations, we have analyzed in~\cite{Actis:2008rb} 
the set of SM diagrams which are suppressed in the effective-theory 
framework by a factor $\delta^{3\slash 2}$ rather than $\delta^2$ with 
respect to the LO cross section, and can be classified as parametrically 
dominant NNLO corrections. They can be conveniently organized
in three sub-sets: 1) mixed hard-Coulomb corrections; 2) interference 
effects of Coulomb and soft (collinear) photons; 3) radiative corrections 
to the Coulomb potential.

\begin{wrapfigure}{l}{0.7\columnwidth}
  \centerline{\includegraphics[width=0.63\columnwidth]{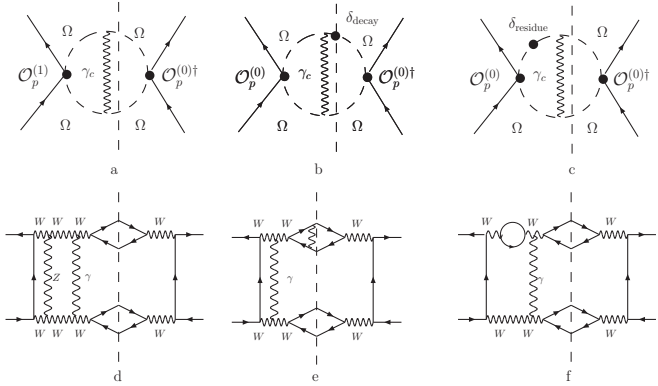}}
  \caption{mixed hard-Coulomb corrections in the effective theory 
           (first line) and in the Standard Model (second line).}
  \label{fig:NLOcount}
\end{wrapfigure}

Mixed hard-Coulomb corrections, given by diagrams with a Coulomb 
photon and one insertion of a hard NLO correction, are illustrated
in Figure~\ref{fig:NLOcount}. Here a hard correction has been 
inserted at the: {\rm a}) production stage, replacing the LO 
production operator ${\cal O}_p^{(0)}$ with the NLO expression 
${\cal O}_p^{(1)}$ as in Figure~\ref{fig:NLOcount}a; {\rm b}) decay stage, 
as graphically shown in Figure~\ref{fig:NLOcount}b by the insertion of 
the black dot labeled $\delta_{\text{decay}}$, summarizing flavor-specific 
contributions to $W$ decay; {\rm c}) propagation stage, as illustrated by 
the $\delta_{\text{residue}}$ insertion in Figure~\ref{fig:NLOcount}c. 
The last contribution is inherent
to the inclusion of wave-function renormalization factors in the 
effective-theory matching coefficients. SM counterparts for all three 
cases are shown in Figure~\ref{fig:NLOcount}d, Figure~\ref{fig:NLOcount}e 
and Figure~\ref{fig:NLOcount}f.

\begin{wrapfigure}{r}{0.7\columnwidth}
  \centerline{\includegraphics[width=0.63\columnwidth]{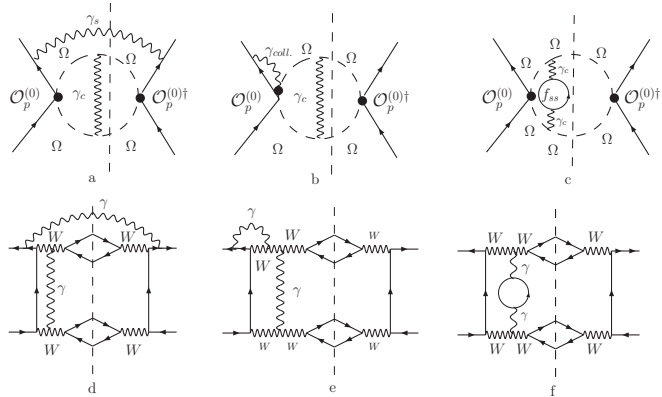}}
  \caption{diagrams involving Coulomb, soft and collinear 
           photons and corrections to the Coulomb potential in the effective 
           theory (first line) and in the Standard Model (second line).}
  \label{fig:NLOcount2}
\end{wrapfigure}

Interference effects of Coulomb and soft ($\gamma_s$) or 
collinear ($\gamma_{coll.}$) photons are shown in 
Figure~\ref{fig:NLOcount2}a and Figure~\ref{fig:NLOcount2}b
with their SM counterparts in Figure~\ref{fig:NLOcount2}d 
and Figure~\ref{fig:NLOcount2}e. As discussed
in~\cite{Actis:2008rb}, they are naturally merged with
the mixed hard-Coulomb corrections at the production stage 
of Figure~\ref{fig:NLOcount}a.
 
Radiative corrections to the Coulomb potential due to
the insertion of a semi-soft fermion bubble ($f_{ss}$) 
are shown in Figure~\ref{fig:NLOcount2}c and 
Figure~\ref{fig:NLOcount2}f.

\section{Results}
\label{sec:num}

The NNLO total cross section follows from the convolution of the
corrections shown in Figure~\ref{fig:NLOcount}
and Figure~\ref{fig:NLOcount2} with the electron structure functions
provided in ~\cite{Skrzypek:1992vk}, in order to re-sum collinear
logarithms from initial-state radiation.

\begin{wraptable}{l}{0.44\columnwidth}\centerline{
\begin{tabular}{|c|c|c|}
    \hline
    $\sqrt{s}$\,[GeV] & $\sigma_{\rm NLO}$ [fb] & $\Delta\sigma_{\rm NNLO}$ [fb] \\
    \hline
    161 &117.81(5)& 0.087 \\\hline
    164 &234.9(1) & 0.544 \\\hline
    167 &328.2(1) & 0.936 \\\hline
    170 &398.0(2) & 1.207 \\\hline
\end{tabular}}
\caption{NLO total cross section for $e^- e^+ \to \mu^- 
  \overline{\nu}_\mu u \overline{d} + X$ and NNLO shift.}
\label{tab:limits}
\end{wraptable}

Results for the NLO evaluation of~\cite{Beneke:2007zg} and
the NNLO shifts of~\cite{Actis:2008rb},
ranging from $0.07 \%$ for $\sqrt{s}= 161$ GeV to 
$0.3 \%$ for $\sqrt{s}= 170$ GeV,
are summarized in Table~\ref{tab:limits}. 
Using the procedure of~\cite{Beneke:2007zg},
we have found that the impact of the dominant NNLO corrections on the $W$-mass
determination is about $3$ MeV. The result is well below the
$6$ MeV error in the measurement from an energy scan
in electron-positron collisions.

We conclude observing that, although a differential calculation in 
the effective theory is not currently feasible (see developments for top-antitop
production in~\cite{Hoang:2008ud}), the analysis of~\cite{Actis:2008rb} 
has shown that the inclusive NNLO result is adequate
for practical applications.

\section{Acknowledgments}

M. Beneke, P. Falgari and C. Schwinn are gratefully acknowledged for the collaboration.
Diagrams have been drawn with {\sc Axodraw}/{\sc Jaxodraw}~\cite{Vermaseren:1994je}.


\begin{footnotesize}

\end{footnotesize}


\end{document}